\documentclass[pre,aps,twocolumn,superscriptaddress,showpacs]{revtex4}
\usepackage[T1]{fontenc}
\usepackage[latin1]{inputenc}
\usepackage{graphicx}
\usepackage{nicefrac}
\usepackage{fancyvrb}
\usepackage{amsmath,amssymb,amstext}
\usepackage{siunitx}
\usepackage{url}
\usepackage{color}

\renewcommand{\vec}[1]{\mathbf{#1}}
\newcommand{\ms}{\color{black}}

\begin{document}

\title{Superadiabatic forces in the dynamics of the
  one-dimensional Gaussian core model}

\author{Elias Bernreuther}
\affiliation{Theoretische Physik II, Physikalisches Institut, 
  Universit{\"a}t Bayreuth, D-95440 Bayreuth, Germany}

\author{Matthias Schmidt}
\affiliation{Theoretische Physik II, Physikalisches Institut, 
  Universit{\"a}t Bayreuth, D-95440 Bayreuth, Germany}
\email{Matthias.Schmidt@uni-bayreuth.de}


\date{27 June 2016}
\begin{abstract}
    Using Brownian dynamics computer simulations we investigate the
    dynamics of the one-body density and one-body current in a
    one-dimensional system of particles that interact with a repulsive
    Gaussian pair potential. We systematically split the internal
    force distribution into an adiabatic part, which originates from
    the equilibrium free energy, and a superadiabatic contribution,
    which is neglected in dynamical density functional theory.  We
    find a strong dependence of the magnitude and phase of the
    superadiabatic force distribution on the initial state of the
    system. While the magnitude of the superadiabatic force is small
    if the system evolves from an equilibrium state inside of a
    parabolic external potential, it is large for particles with
    equidistant initial separations at high temperature. We analyze
    these findings in the light of the known mean-field behavior of
    Gaussian core particles and discuss a multi-occupancy
    mechanism which generates superadiabatic forces that are out of
    phase with respect to the adiabatic force.
\end{abstract}
\maketitle

\section{Introduction}
\label{chap:intro}

Approximating the time evolution of a physical system as an adiabatic process, which proceeds through a series of equilibrium states, has been a helpful approach both in classical physics and in quantum mechanics \cite{Born1928}. In dynamical density functional theory (DDFT) \cite{evans79,marinibettolomarconi99,dzubiella03,archer04ddft,Hansen13,royallSedimentation,evans16pc}, which is a framework for describing the nonequilibrium evolution of classical many-body systems, such an adiabatic approximation is employed on the level of the position-dependent one-particle density distribution. Thereby the correlations between particle positions are approximated by those of a fictitious equilibrium system with the same one-body density profile as the  nonequilibrium system of interest \cite{Hansen13}. This substitution implies the assumption that the relaxation time of particle correlations is short compared to the time scale of the processes of interest. However, this assumption is not justified, e.g., for strongly confined systems, which leads to a failure of DDFT in those cases \cite{fortini14prl}. In order to accurately describe the dynamics of such systems, superadiabatic forces need to be included. In recent work Fortini \emph{et al.}\ \cite{fortini14prl} have proposed a computational scheme for systematically dividing internal forces in a system of Brownian particles into an adiabatic and a superadiabatic contribution. By applying their method to a system of confined hard rods in one dimension,  which is a common model system \cite{penna06,reinhardt12} for studying dynamical features, these authors found that the superadiabatic part of the force in that system is not a small correction and cannot easily be related to the adiabatic part. Furthermore the results depended sensitively on the chosen initial condition. The behavior of these superadiabatic forces is at present not well understood.\par
In order to gain more insight into the splitting into adiabatic and superadiabatic forces we apply the simulation scheme of Fortini \emph{et al.}\ to a one-dimensional system that is qualitatively different from hard rods. We choose the pair potential of the Gaussian core model (GCM) \cite{stillinger76}, which  constitutes an approximation to the effective interactions of polymer coils. We investigate the superadiabatic forces and their temperature dependence. We start either from an equilibrium initial state, for which the strength of an external parabolic potential is instantaneously altered in order to drive the time evolution, or from a range of nonequilibrium initial states, in which the particles are located with equidistant separations. In both cases we apply a parabolic external potential for confinement.

Our motivation for studying a one-dimensional system stems from similar
studies of equilibrium systems, for, e.g., colloid-polymer mixtures
\cite{brader01oned} or non-additive hard core mixtures
\cite{santos07oned}, where exact solutions are possible in
one-dimension.  Experimentally, single-file diffusion has been
demonstrated for colloids confined in one-dimensional channels
\cite{lutz04}.
{\ms In the system of one-dimensional hard rods Penna and Tarazona
  [10] investigated the dynamic decay of (imposed) periodic density
  oscillations and compared DDFT results with Brownian Dynamics (BD)
  simulations. The authors developed a theory for the dynamics of the
  two-body density, based on the equilibrium factorization of the
  three-body density, that yields results superior to those from DDFT.  }

For the GCM we find that the magnitude of the superadiabatic contribution to the total force depends sensitively on the initial state of the system. This echoes the observations for the system of hard rods, where this dependence was also found to be strong \cite{fortini14prl}. If the time evolution starts from an equilibrium state, the adiabatic approximation remains accurate during the nonequilibrium time evolution, which confirms the validity of DDFT in this situation. However, for equidistant initial particle locations we find large superadiabatic forces, which are out-of-phase with respect to the adiabatic force and tend to grow with temperature. We identify the occupation of density peaks, which in reality correspond to individual particles, by multiple particles \cite{fortini14prl} to be one of the reasons for the occurrence of these superadiabatic forces.

{\ms Our results stem from explicit many-body BD and Monte Carlo (MC)
  computer simulations. We leave the construction of corresponding
  theoretical approximations to future work, based on, e.g., the
  approach of  Ref.~\cite{penna06} or on power functional theory
  \cite{schmidt13pft} }

The paper is organized as follows. In Sec.~\ref{chap:theory} we lay
out the theoretical background, including the underlying Brownian
many-body dynamics (Sec.~\ref{sec:many-bodyDynamics}), the reduced
one-body description (Sec.~\ref{sec:one-bodyDescription}), and details
about the definition and the equilibrium density functional
description of the GCM (Sec.~\ref{sec:gcm}). In
Sec.~\ref{chap:methods} we describe the simulation methods that we
use, including an overview of the units and initial states
(Sec.~\ref{sec:unitsEtc}), BD in nonequilibrium for obtaining the
full time evolution (Sec.~\ref{sec:bd}), and the MC method and
adiabatic iteration scheme for constructing the adiabatic state
(Sec.~\ref{sec:mc}).  We present our results in Sec.~\ref{chap:results},
including those for initially equilibrated
(Sec.~\ref{sec:equilibrated}) and equidistant
(Sec.~\ref{sec:equidistant}) states. We conclude in
Sec.~\ref{chap:concl}.

\section{Theoretical background}
\label{chap:theory}

\subsection{Many-body dynamics}
\label{sec:many-bodyDynamics}

We consider a system of $N$ interacting Brownian particles. In an overdamped system with only potential forces the positions $\vec{r}_i$ ($i=1,\ldots,N$) of the Brownian particles satisfy the Langevin equation \cite{Hansen13}
\begin{equation}
\label{eq:langevin}
\xi\dot{\vec{r}}_i(t)=-\nabla_i U_{\rm tot}(\vec{r}^N,t)+\vec{R}_i(t)
\end{equation}
where $\xi$ is the friction coefficient, the dot indicates the total derivative with respect to time $t$, $\nabla_i$ is the partial derivative with respect to $\vec{r}_i$, and $U_{\rm tot}(\vec{r}^N,t)$ indicates the total potential energy, which depends on the positions of all $N$ particles. The vector $\vec{R}_i(t)$ represents a temperature-dependent, randomly fluctuating force, which is due to collisions of the Brownian particles with (implicit) particles of the solvent, which take place on a far shorter time-scale than the relaxation of the initial velocity of the Brownian particles. The random forces on two distinct particles are uncorrelated and the correlation time for forces that act on the same particle is infinitesimally short, therefore the dyadic product is on average
\begin{equation}
\label{eq:noisecorrelator}
{\langle\vec{R}_i(t)\vec{R}_j(t^{\prime})\rangle}=2\pi\mathbb{I} { R}_0\delta_{ij}\delta(t-t^{\prime}),
\end{equation}
where $\mathbb{I}$ indicates the $3\times 3$ unit matrix, $R_0 = 3\xi
k_\mathrm{B}T/\pi$ measures the strength of the fluctuating force,
$\delta_{ij}$ indicates the Kronecker delta, and $i,j=1\ldots N$; here
$k_\mathrm{B}$ is the Boltzmann constant and $T$ is absolute temperature.

As an alternative to considering the dynamics of individual particles, the system can be characterized via the $N$-particle probability density $\psi(\vec{r}^N,t)$ for observing a configuration $\vec{r}^N=\{\vec{r}_1,\ldots,\vec{r}_N\}$ of positions of all $N$ particles at time $t$ (irrespective of their momenta). The time evolution of $\psi(\vec{r}^N,t)$ is described by the Smoluchowski equation \cite{Hansen13}:
\begin{equation}
\label{eq:smoluchowski}
\xi\frac{\partial \psi(\vec{r}^N,t)}{\partial t}=\sum_{i=1}^{N}\nabla_i \cdot [k_\mathrm{B}T\nabla_i + \nabla_iU_{\rm tot}(\vec{r}^N,t)]\psi(\vec{r}^N,t).
\end{equation}
If the internal interactions stem from a pair potential $\phi$, then the total potential energy reads
\begin{equation}
\label{eq:potential}
U_{\rm tot}(\vec{r}^N,t)=\sum_{i}\sum_{j<i}\phi(|\vec{r}_i-\vec{r}_j|)+\sum_{i}V_{\mathrm{ext}}(\vec{r}_i,t)
\end{equation}
where $V_{\mathrm{ext}}(\vec{r},t)$ is the external potential, which in general depends on position $\vec{r}$ and time $t$.

\subsection{One-body description}
\label{sec:one-bodyDescription}
The time-dependent one-body density $\rho(\vec{r},t)$ and the two-body density $\rho^{(2)}(\vec{r},\vec{r}^\prime,t)$ are obtained by integrating $\psi(\vec{r}^N,t)$ over all but one and all but two particle coordinates, respectively, and multiplying by the appropriate normalizing factor \cite{fortini14prl},
 \begin{align}
 \rho(\vec{r}_1,t)&=N \int \mathrm{d}\vec{r}_2 ...\!\int \mathrm{d}\vec{r}_N \,\ \psi(\vec{r}^N,t),\\
 \rho^{(2)}(\vec{r}_1,\vec{r}_2,t)&=N(N-1) \int \mathrm{d}\vec{r}_3 ...\!\int \mathrm{d}\vec{r}_N \,\ \psi(\vec{r}^N,t).
 \end{align}
Consequently, by integrating both sides of (\ref{eq:smoluchowski}) over $N-1$ particle coordinates one obtains an evolution equation for the one-body density, as shown by Archer and Evans \cite{archer04ddft}. These authors also included three- and higher-body interactions, which are not of interest for the current work. We will therefore restrict ourselves to pair interactions of the form (\ref{eq:potential}).
As a result of the integration one obtains the continuity equation,
\begin{equation}
\frac{\partial\rho(\vec{r},t)}{\partial t}=-\nabla\cdot\vec{J}(\vec{r},t),
\end{equation}
where the one-body current,
\begin{equation}
\label{eq:current}
\vec{J}(\vec{r},t)=\xi^{-1}\rho(\vec{r},t)\vec{F}(\vec{r},t),
\end{equation}
is due to the one-body force,
\begin{equation}
\label{eq:force}
\vec{F}(\vec{r},t)=-k_\mathrm{B}T{\nabla\mathrm{ln}\rho(\vec{r},t)}-\nabla V_{\mathrm{ext}}(\vec{r},t)+\vec{F}_{\mathrm{int}}(\vec{r},t),
\end{equation}
being the sum of an ideal diffusion part, the external force, and the internal force,
$\vec{F}_\mathrm{int}(\vec{r},t)$.
For convenience, we define the internal force density $\vec{I}_\mathrm{int}(\vec{r},t)$ via the internal force and the one-body density as
\begin{equation}
\label{eq:forcedensity}
\vec{I}_\mathrm{int}(\vec{r},t) \equiv
  \vec{F}_\mathrm{int}(\vec{r},t)\rho(\vec{r},t).
\end{equation}
For the present case of internal pair interactions, the internal force density can be expressed as the integral
\begin{equation}
\label{eq:intforcedensity}
\vec{I}_\mathrm{int}(\vec{r},t)=-\int \mathrm{d}\vec{r}^{\prime} \rho^{(2)}(\vec{r},\vec{r}^{\prime},t)\nabla\phi(|\vec{r}-\vec{r}^{\prime}|).
\end{equation}
Archer and Evans derived the DDFT equation of motion by approximating
$\rho^{(2)}(\vec{r},\vec{r}^{\prime},t)$ as an ``adiabatic'' two-body
density
\begin{equation}
\label{eq:adiabaticApproximation}
  \rho^{(2)}(\vec{r},\vec{r'},t)=
  \rho^{(2)}_{\mathrm{ad},t}(\vec{r},\vec{r}^{\prime}),
\end{equation}
which is defined as the equilibrium two-body density of a system with
the same internal interactions, but with an external potential
$V_{\mathrm{ad},t}(\vec{r})$ that is adjusted such that the
corresponding equilibrium one-body density,
$\rho_{\mathrm{ad},t}(\vec{r})$, equals the instantaneous one-body
density in the nonequilibrium system at time~$t$:
\begin{equation}
\label{eq:addensity}
\rho_{\mathrm{ad},t}(\vec{r})=\rho(\vec{r},t).
\end{equation} Note that $\rho^{(2)}_{\mathrm{ad},t}(\vec{r},\vec{r}^{\prime})$ does not depend on time in the given adiabatic system, as this is in equilibrium. However, the adiabatic system itself changes with time as $\rho(\vec{r},t)$ changes. As $\rho^{(2)}_{\mathrm{ad},t}(\vec{r},\vec{r}^{\prime})$ is an equilibrium density, Archer and Evans then proceed using the sum rule \cite{archer04ddft}
\begin{equation}
\label{eq:sumrule}
-k_\mathrm{B}T\rho_{{\rm ad},t}(\vec{r})\nabla
c^{(1)}(\vec{r})=\int\mathrm{d}\vec{r}^{\prime}\rho^{(2)}_{\mathrm{ad},t}(\vec{r},\vec{r^{\prime}})\nabla\phi(|\vec{r}-\vec{r}^{\prime}|),
\end{equation}
which relates the internal force density in equilibrium to the
one-body direct correlation function $c^{(1)}(\vec{r})$ in the
 adiabatic system. Here the direct correlation function is defined
in the grand canonical ensemble. Very recently, de las Heras et
al.\ \cite{delasheras16pc} argued that this is at odds with the
$N$-particle character of the dynamics. These authors propose a
consistent treatment on the basis of the equilibrium canonical
decomposition scheme of Ref.~\cite{delasheras14prl} applied to the
adiabatic state. This approach leads to a particle-conserving
dynamical theory \cite{delasheras16pc}.

The conventional derivation involves substituting the identity
\begin{equation}
\label{eq:c1fex}
-k_\mathrm{B}T c^{(1)}(\vec{r})=
\frac{\delta F^{\mathrm{ex}}[{\rho}]}{\delta\rho(\vec{r})}
\end{equation}
from equilibrium density functional theory, where $F^{\mathrm{ex}}[{\rho}]$ is the excess (over ideal gas) Helmholtz free energy functional, in order to obtain the DDFT equation of motion~\cite{archer04ddft}
\begin{equation}
\xi\frac{\partial\rho(\vec{r},t)}{\partial t}= \nabla\cdot\left(\rho(\vec{r},t)\nabla\left.\frac{\delta F^{\mathrm{tot}}[{\rho_{}}]}{\delta\rho_{\mathrm{}}(\vec{r})} \right\vert_{\rho_{}(\vec{r})=\rho(\vec{r},t)}\right),
\end{equation}
where $F^\mathrm{tot}[\rho]=F^\mathrm{exc}[\rho]+ k_{\rm B}T\int
\mathrm{d}\vec{r} \rho(\vec{r})[\ln(\rho(\vec{r})\Lambda^3)-1]+ \int
\mathrm{d}\vec{r} {V_{\rm ext}(\vec{r},t)}\rho(\vec{r})$ is the
total Helmholtz free energy functional (with the irrelevant thermal
wavelength $\Lambda$).

More recently, however, Fortini \emph{et al.}\ \cite{fortini14prl} showed by
using BD and MC simulations that the adiabatic
approximation \eqref{eq:adiabaticApproximation}, even when performed consistently with $N$ particles,
 produces qualitatively wrong results for a system of hard rods in one dimension. Their work was motivated by recent theoretical progress, where superadiabatic forces arise from the functional derivative of an excess power dissipation functional \cite{schmidt13pft}. Hence, following Fortini \emph{et al.}\ adiabatic and superadiabatic forces are equally valid contributions to the internal force density \eqref{eq:intforcedensity}, which can explicitly be written as the sum of the adiabatic force density, defined as
\begin{equation}
\label{eq:defiad}
\vec{I}_{\mathrm{ad},t}(\vec{r})=
-\int\mathrm{d}\vec{r}^{\prime}\rho^{(2)}_{\mathrm{ad},t}(\vec{r},\vec{r^{\prime}})\nabla\phi(|\vec{r}-\vec{r}^{\prime}|)
\end{equation}
and the superadiabatic force density, defined as
\begin{equation}
\label{eq:defisup}
\vec{I}_{\mathrm{sup}}(\vec{r},t) \equiv \vec{I}_\mathrm{int}(\vec{r},t)-\vec{I}_{\mathrm{ad},t}(\vec{r}).
\end{equation}
where $\vec{I}_\mathrm{int}(\vec{r},t)$ is given by (\ref{eq:forcedensity}) or (\ref{eq:intforcedensity}).

\subsection{Gaussian core model and the mean-field approximation}
\label{sec:gcm}
The one-component GCM is a system with pairwise interactions between particles, where the pair potential as a function of center-center distance $r$ is given by
\begin{equation}
\label{eq:gcm}
\phi(r)=\epsilon \exp\left(-r^2/\sigma^2\right),
\end{equation}
where the constant $\epsilon$ is the potential energy at zero separation of the two particles and the constant $\sigma$ determines the length scale. The GCM was first introduced by Stillinger in 1976 \cite{stillinger76} and is regarded as a good approximation for the effective interaction of centers of mass of polymer coils \cite{Hansen13}. In particular, for two isolated non-intersecting chains, $\epsilon$ is of order $2k_\mathrm{B}T$ and $\sigma$ is approximately equal to the radius of gyration of the polymer coils \cite{louis00}.
In equilibrium the three-dimensional GCM is known to behave like a mean-field fluid over a broad range of densities and temperatures \cite{louis00} and the mean-field free energy density functional \cite{archer01,archer02,archer02el,archer03,archer04},
 \begin{equation}
 \label{eq:meanfieldfunctional}
F^{\mathrm{ex}}[ \rho]=\frac{1}{2}\int\mathrm{d}\vec{r}\int\mathrm{d}\vec{r}^{\prime}\rho(\vec{r})\phi(|\vec{r}^{\prime}-\vec{r}|)\rho(\vec{r}^{\prime}),
 \end{equation}
 provides a good approximation.
 Substituting the functional (\ref{eq:meanfieldfunctional}) into (\ref{eq:c1fex}) and using (\ref{eq:sumrule}) and the definition of the adiabatic force density $\vec{I}_{\mathrm{ad},t}(\vec{r})$, (\ref{eq:defiad}), yields
\begin{equation}
\label{eq:iadmeanfield}
\vec{I}_{\mathrm{ad},t}(\vec{r})=
-\rho_{{\rm ad},t}(\vec{r})
\nabla\int\mathrm{d}\vec{r}^{\prime}\phi(|\vec{r}^{\prime}-\vec{r}|)
\rho_{{\rm ad},t}(\vec{r}^{\prime}).
\end{equation}
Comparing (\ref{eq:iadmeanfield}) to (\ref{eq:defiad}) shows that the two-body density has been factorized into a product of one-body densities:
\begin{equation}
\label{eq:mffactorization}
\rho^{(2)}_{\mathrm{ad},t}(\vec{r},\vec{r}^{\prime})=
\rho_{{\rm ad},t}(\vec{r})\rho_{{\rm ad},t}(\vec{r}^{\prime}).
\end{equation}
{\ms For obtaining the results presented below, we do not employ such
  approximations, but rather solve the full problem(s) using both
  explicit BD and (Metropolis) MC many-body computer simulations
  \cite{Hansen13}, which we outline in the following.}

\section{Simulation methods}
\label{chap:methods}

\subsection{Units, physical setup and initial states}
\label{sec:unitsEtc}

For our simulation work we chose $\sigma$, $\epsilon$, and $\xi$ as the
fundamental units. Time is then given in units of
$\tau_0=\sigma^2\xi/\epsilon$, which is a time scale that
characterizes the dynamics due to the internal interactions. We indicate
the reduced temperature by $T^*= k_\mathrm{B}T/\epsilon$. We
consider $N=10$ particles with position coordinates $x_i$, $i=1\ldots
10$, in one dimension
inside of a parabolic external potential
\begin{equation}
\label{eq:vext}
V_{\mathrm{ext}}(x)=\frac{k}{2}x^2,
\end{equation}
where $x$ is the space coordinate and the parameter $k$ determines the curvature of the parabola, and hence the strength of confinement. We investigate the following two different classes of initial conditions.

(i) {\it Equilibrium initial states.} Here the system is equilibrated
in the external potential \eqref{eq:vext} with $k=k_0$ for times
$-5\tau_0\leq t\leq 0$. For $t>0$ the value of $k$ is then
set to $k_1<k_0$, which clearly changes the equilibrium state in the
external potential. We consider the reduced temperatures $T^*=0.02$,
0.1, and 0.5.

(ii) {\it Equidistant initial states.} At $t=0$ the particles are
placed equidistantly and symmetrically around the origin with a
nearest-neighbor distance $d$. For two values of the reduced
temperature $T^*=0.1, 0.5$ we consider the distance ratios
$d/\sigma=0$, 0.11, 0.22, 0.56, 0.83, 1.11, 1.67, and 2.22.

In both cases (i) and (ii) the system subsequently evolves until the time $t=t_\mathrm{s}$ where the instantaneous one-body density $\rho(x,t_\mathrm{s})$ and two-body density $\rho^{(2)}(x,x^\prime,t_\mathrm{s})$ as well as the one-body current $J(x,t_\mathrm{s})$ are sampled. The instantaneous internal force $F_\mathrm{int}(x,t_\mathrm{s})$ and internal force density $I_{\rm int}(x,t_\mathrm{s})$ are then obtained by numerically evaluating the integral (\ref{eq:intforcedensity}). {\ms Here and throughout our one-dimensional study we indicate the
   vectorial quantities, such as the force density and the particle
   current, as (effective) scalars in the notation.} Furthermore, the knowledge of $\rho(x,t_\mathrm{s})$ allows us to identify the corresponding adiabatic potential $V_{\mathrm{ad},t}(x)$ and simulate the adiabatic system (as defined in Sec.~\ref{sec:one-bodyDescription}). Thus, we can obtain the adiabatic and superadiabatic contributions to the internal force and force density. The concrete implementation of the nonequilibrium and the equilibrium simulation as well as the scheme for identifying the adiabatic potential are described in the following.

\subsection{Brownian dynamics in nonequilibrium}
\label{sec:bd}

Following Ref.~\cite{fortini14prl} we use the BD method to simulate the nonequilibrium dynamics.
The BD simulation consists of numerical integration of the overdamped Langevin equation of motion (\ref{eq:langevin}) in discrete time steps
$\Delta t=0.0005\tau_0$. The system was advanced via
the Euler algorithm.
In order to perform the average
of the stochastic differential equation (\ref{eq:langevin}),
$10^5$--$10^7$ independent trajectories were generated for each case
considered. At each time step the random forces $\vec{R}_i(t)$
are generated according to a Gaussian probability distribution with
standard deviation
$\sqrt{2 \Delta t k_\mathrm{B}T \xi}=\sqrt{2\Delta t^* T^*}\tau_0\epsilon/\sigma$,
where $\Delta t^*=\Delta t/\tau_0$ is the scaled time step.
We use one-
and two-dimensional histograms for obtaining results for the one- and two-body quantities, respectively, with numbers of bins in each
dimension between $300$ and $1600$ and bin sizes
ranging from $0.02\sigma$ to $0.1\sigma$.

\subsection{Monte Carlo and adiabatic iteration scheme}
\label{sec:mc}

In order to find the adiabatic external potential, for which the
density in the adiabatic system, $\rho_{\mathrm{ad},t_\mathrm{s}}(x)$,
is equal to $\rho(x,t_\mathrm{s})$ [cf.~(\ref{eq:addensity})], we
employ an iterative scheme of successive MC simulations.
Here $\rho(x,t_\mathrm{s})$ is sampled in the nonequilibrium BD
simulation as described above. {\ms Our MC scheme is a standard
  many-body (Metropolis) method, where we use single-particle moves in
  order to update the particle positions, and hence obtain equilibrium
  averages.}  As the GCM can in many equilibrium situations be well
approximated by a mean-field approach \cite{louis00}, we find it
useful to then obtain the initial value of the adiabatic external
potential from the mean-field approximation,
\begin{align}
V_{\mathrm{ad},t_\mathrm{s}}^{(0)}(x)&=-k_\mathrm{B}T\ln(\rho(x,t_\mathrm{s})\Lambda)
-\int\mathrm{d}x^\prime{\rho(x^\prime,t_\mathrm{s})\phi(|x^\prime-x|).}
\end{align}
Here the thermal wavelength, $\Lambda$, generates only an additive constant to the adiabatic external potential and can therefore be neglected.
For the iteration the adiabatic potential is discretized to one value per histogram bin. In each iteration step we run an MC simulation under the influence of the adiabatic external potential and compare the resulting density $\rho_{\mathrm{ad},t_{\rm s}}(x)$ in each bin to $\rho(x,t_\mathrm{s})$, as known from BD. If $\rho_{\mathrm{ad},t_{\rm s}}(x)>\rho(x,t_\mathrm{s})$, then we increase $V_{\mathrm{ad},t_{\rm s}}(x)$ in the respective bin; if $\rho_{\mathrm{ad},t_{\rm s}}(x)<\rho(x,t_\mathrm{s})$, then we decrease $V_{\mathrm{ad},t_{\rm s}}(x)$. In our particular implementation the potential in the $(i+1)$th step follows from the result of the $i$th step according to
\begin{equation}
V_{\mathrm{ad},t_{\rm s}}^{(i+1)}(x)=V_{\mathrm{ad},t_{\rm s}}^{(i)}(x)+\alpha\ln\left(\frac{\rho^{(i)}_{\mathrm{ad},t_{\rm s}}(x)}{\rho(x,t_\mathrm{s})}\right)
\end{equation}
with $\alpha$ as a free parameter, typically chosen as $\alpha=1$ or $\alpha=0.5$. Here the superscripts indicate the iteration steps. We continue the iteration until the cutoff criterion $|\rho_{\mathrm{ad},t_{\rm s}}(x)-\rho(x,t_\mathrm{s})|<0.01\rho(x,t_\mathrm{s})$ is reached. Each MC run consists of up to $2\cdot10^9$ MC steps (attempted particle updates), as necessary to obtain adequate statistical quality. The acceptance rate of attempted single particle moves falls within a range between $20\%$ and $50\%$.  The adiabatic two-body density $\rho^{(2)}_{\mathrm{ad},t_{\rm s}}(x,x')$ is sampled from a separate MC run in the final adiabatic potential.
The adiabatic and superadiabatic force densities and forces are then calculated from (\ref{eq:defiad}) and (\ref{eq:defisup}), respectively.

\section{Results}
\label{chap:results}

\subsection{Equilibrated initial states}
\label{sec:equilibrated}

\begin{figure}
  \includegraphics[width=\columnwidth]{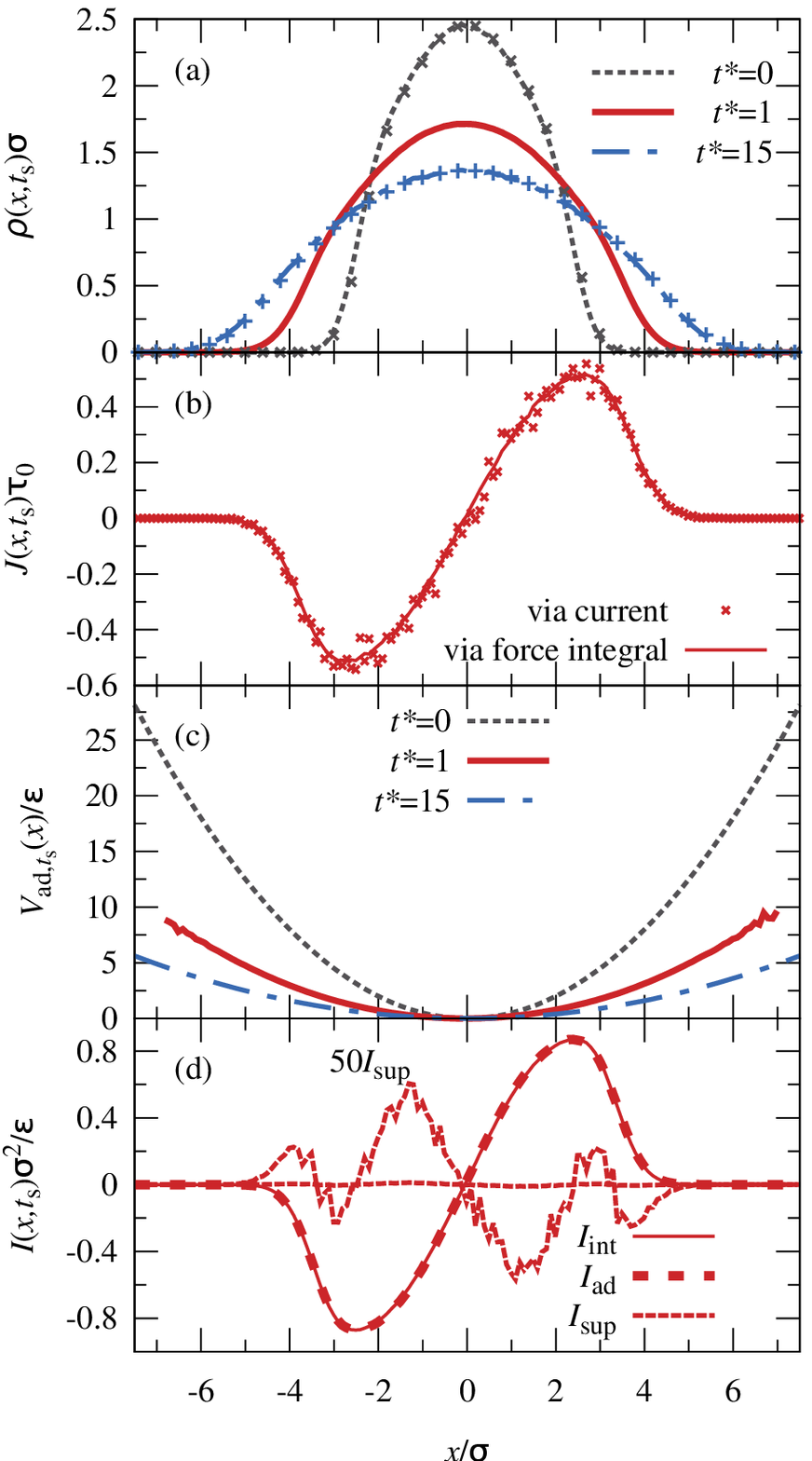}
  \caption{\label{fig:overviewequi} Overview of the one-body profiles as a function of the scaled distance $x/\sigma$ in an initially equilibrated system at reduced temperature $T^*=0.5$. (a) Time evolution of the density $\rho(x,t)$ at times $t^*\equiv t/\tau_0=0$ (dashed line), $1$ (solid line) and $15$ (long-short dashed line) obtained from BD (lines). For comparison the equilibrium densities $\rho(x)$ obtained from MC (symbols) are shown for $k=\epsilon/\sigma^2$ and $0.2\epsilon/\sigma^2$. (b) Comparison of the scaled current $J(x,t_\mathrm{s})\tau_0$ from direct sampling (symbols) and from evaluation of the force integral (\ref{eq:current}) (line). (c) Adiabatic external potential $V_{\mathrm{ad},t_{\rm s}}(x)$ in units of $\epsilon=2k_\mathrm{B}T$. (d) Total internal, $I_\mathrm{int}(x,t_{\rm s})$, adiabatic, $I_{\mathrm{ad},t_\mathrm{s}}(x)$, and superadiabatic, $ I_\mathrm{sup}(x,t_{\rm s})$, force density in units of $\epsilon/\sigma^2$ at $t^*=1$. $I_\mathrm{sup}(x,t_{\rm s})$ is additionally shown multiplied by a factor of $50$ (see label).}
\end{figure}

We first consider a system of $N=10$ particles confined inside the
parabolic external potential (\ref{eq:vext}) where
$k_0=\epsilon/\sigma^2$ and $k_1=0.2\epsilon/\sigma^2$ at reduced
  temperature $T^*=0.5$.
As can be seen from Fig.~\ref{fig:overviewequi}(a), the equilibrium density profile $\rho(x,t=0)$ resembles the Gaussian distribution that an ideal gas would produce in the same external potential. However, the actual width is larger than in the ideal gas case due to the repulsion between GCM particles. At $t=15\tau_0$ the density profile has spread outward and relaxed towards the new equilibrium state for $k=k_1$. For both (initial and final) equilibrium densities the results obtained from BD are in very good agreement with MC simulations in the respective external potential.\par
For investigating the nonequilibrium quantities of interest, and in particular the superadiabatic force, we chose the sampling time to be $t_s=\tau_0$. At this time the outward movement of the density profile is reflected by the sign of the current $J(x,t_s=\tau_0)$, which is negative for $x<0$ and positive for $x>0$; see Fig.~\ref{fig:overviewequi}(b). Furthermore, we have checked that results for the current, as sampled directly from the particle velocities $\dot x_i(t_\mathrm{s})$, agree very well with the results for the current calculated from $\rho(x,t_\mathrm{s})$ and $\rho^{(2)}(x,x^\prime,t_\mathrm{s})$ via the integral (\ref{eq:current}). However, the direct sampling produces a higher statistical uncertainty, which we attribute to the presence of the random force in (\ref{eq:langevin}).\par
The result for the adiabatic potential, as shown in Fig.~\ref{fig:overviewequi}(c), is in this case similar to a parabola with curvature $k=0.39\epsilon/\sigma^2$. Finally Fig.~\ref{fig:overviewequi}(d) shows the total internal force density $I_\mathrm{int}(x,t_s)$ as well as its two constituents, the adiabatic force density $I_{\mathrm{ad},t_s}(x)$ and the superadiabatic force density $I_\mathrm{sup}(x,t_s)$. Considering force densities offers the advantage over forces that the former feature a smaller statistical error in regions  where the density is small, i.e., in the present case at the edges of the spatial range that we consider. However, due to the simple relation (\ref{eq:forcedensity}) all conclusions about the superadiabatic behavior of the system remain valid when considering forces. For the case $T^*=0.5$, the total internal force density and the adiabatic force density nearly coincide, while $I_\mathrm{sup}(x,t_\mathrm{s})$ constitutes a small correction, which is roughly two orders of magnitude smaller than $I_\mathrm{int}(x,t_\mathrm{s})$ and $I_{\mathrm{ad},t_{\rm s}}(x)$. This finding suggests that the equilibrium structure of the fluid is preserved during the particular nonequilibrium time evolution that we consider here. The reason for this behavior can be connected to the fact that for the GCM mean-field methods provide a good approximation in a wide range of equilibrium situations \cite{louis00}, which includes the adiabatic system considered here. As mentioned above in Sec.~\ref{sec:gcm}, in the mean-field approximation $\rho^{(2)}(x,x^\prime)$ is factorized into a product of one-body densities (\ref{eq:mffactorization}). This implies that if the GCM also has a mean-field-like structure out of equilibrium, $I_\mathrm{int}(x,t_\mathrm{s})$ and $I_{\mathrm{ad},t_s}(x)$, which are calculated via (\ref{eq:intforcedensity}), must be equal, since $\rho(x,t_s)=\rho_{\mathrm{ad},t_s}(x)$ by construction.

\begin{figure}
  \includegraphics[width=\columnwidth]{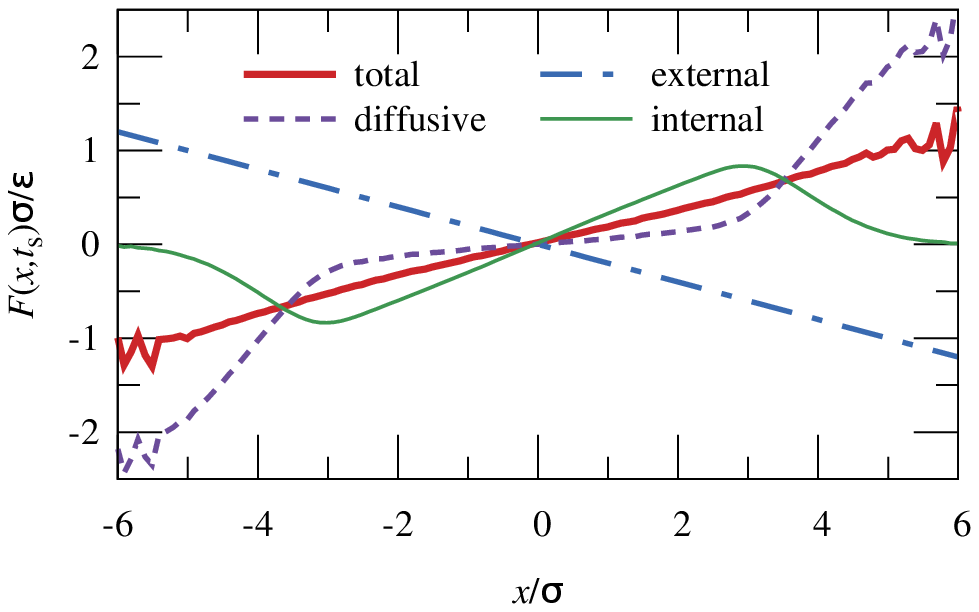}
  \caption{\label{fig:forcesT05}Total force $F(x,t_\mathrm{s})$, diffusive force $-k_\mathrm{B}T\nabla\ln\rho(x,t_\mathrm{s})$, external force $F_\mathrm{ext}(x,t_\mathrm{s})$, and internal force $F_\mathrm{int}(x,t_\mathrm{s})$ for $T^*=0.5$ at sampling time  $t_\mathrm{s}=\tau_0$ in an initially equilibrated system. The statistical noise in the diffusive force grows for large values of $|x|$ as $\rho(x,t_\mathrm{s})$ approaches zero; cf.~Fig.~\ref{fig:overviewequi}.}
\end{figure}

In order to clarify the different physical effects that drive the dynamics, in Fig.~\ref{fig:forcesT05} we show the total force $F(x,t_\mathrm{s})$ for $T^*=0.5$, together with its three constituents: according to (\ref{eq:force}) these are the diffusive force, the external force and the internal force, which we all find to vary linearly with $x$ in the range $-3\sigma<x<3\sigma$ in this particular case. Outside of this range the diffusive force and the external force strongly dominate, while the internal force becomes less relevant.

\begin{figure}
  \includegraphics[width=\columnwidth]{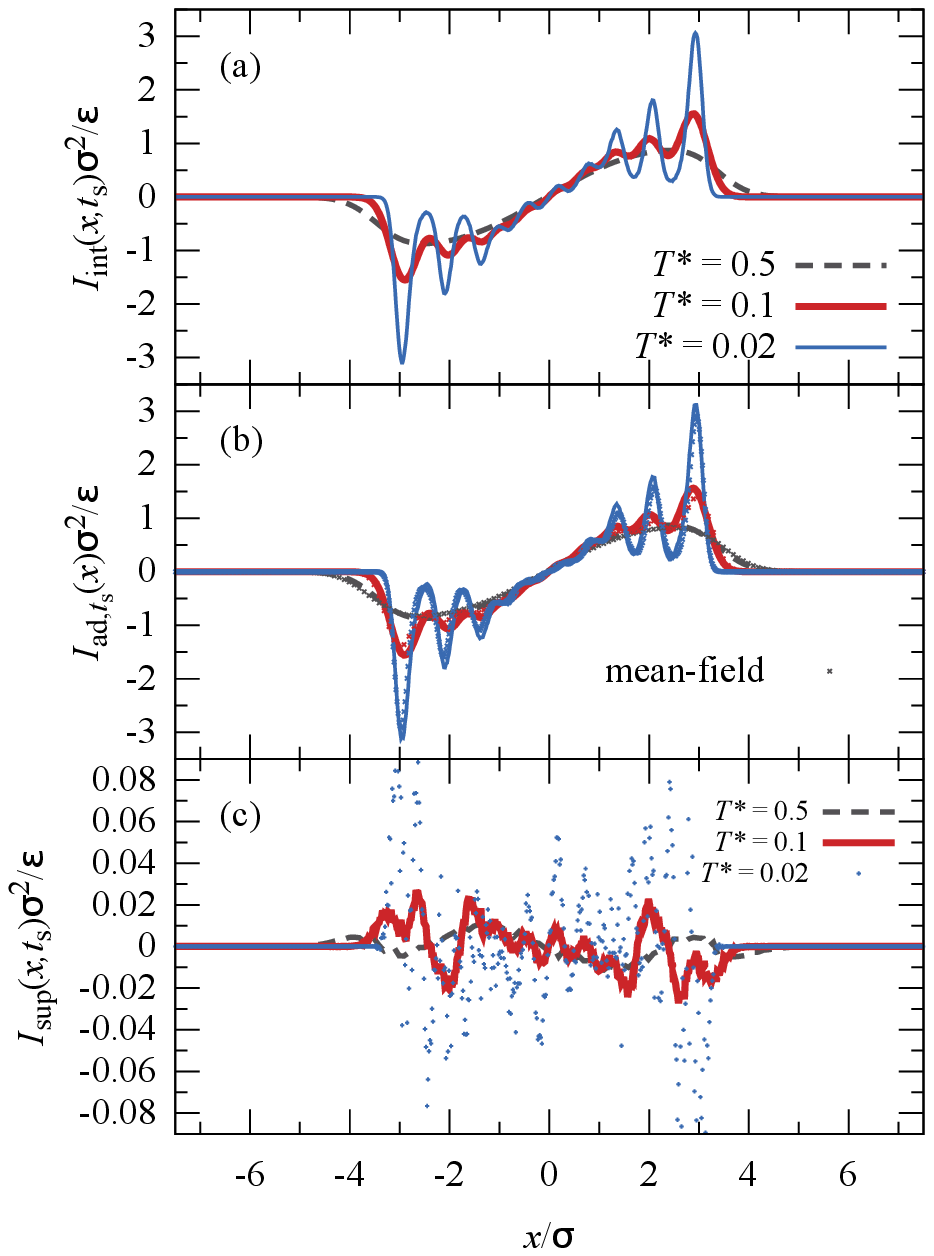}
  \caption{\label{fig:equiI0501002}Comparison of (a) total internal force density $I_\mathrm{int}(x,t_{\rm s})$, (b) adiabatic force density $I_{\mathrm{ad},t_{\rm s}}(x)$ and (c) superadiabatic force density $I_\mathrm{sup}(x,t_{\rm s})$ in units of $\epsilon/\sigma^2$ for scaled temperatures $T^*=0.5$, $0.1$, and $0.02$ (as indicated) at sampling time $t_\mathrm{s}=\tau_0$ for a system that is initially in equilibrium. In (b) the adiabatic force densities $I_{\mathrm{ad},t_{\rm s}}(x)$ (lines) are also compared to the respective mean-field force density (symbols). Note that the scale on the vertical axis is different in~(c).}
\end{figure}

The temperature dependence of the force densities is illustrated in
Fig.~\ref{fig:equiI0501002}, where we show results for the
  internal force density (a), and its adiabatic (b) and superadiabatic
  (c) contributions. 
In addition, in Fig.~\ref{fig:equiI0501002}(b) we compare the adiabatic force density $I_{\mathrm{ad},t_\mathrm{s}}(x)$ to the corresponding mean-field force density, which is obtained from $\rho(x,t_\mathrm{s})$ via (\ref{eq:iadmeanfield}). We observe that, first, the mean-field force density provides a very good approximation to $I_{\mathrm{ad},t_{\rm s}}(x)$. Second, the small deviations from the mean-field force density are of the {same order of magnitude as} the superadiabatic force density $I_\mathrm{sup}(x,t_\mathrm{s})$. This holds also for the cases $T^*=0.1$ and $T^*=0.02$, where the force density becomes more structured and develops oscillations on top of its global slope due to reduced overlap between the particles.  These findings support the conjecture that a dynamical mean-field behavior \cite{dzubiella03} is responsible for the agreement of $I_\mathrm{int}(x,t_{\rm s})$ and $I_{\mathrm{ad},t_{\rm s}}(x)$. Generally, the superadiabatic contribution grows as $T^*$ decreases, but remains small for all temperatures that we investigated; see Fig.~\ref{fig:equiI0501002}(c). We conclude that for the special case of a time evolution between two equilibrium states the adiabatic approximation gives good results under all conditions that we considered.

\subsection{Equidistant initial states}
\label{sec:equidistant}

\begin{figure}
    \includegraphics[width=\textwidth]{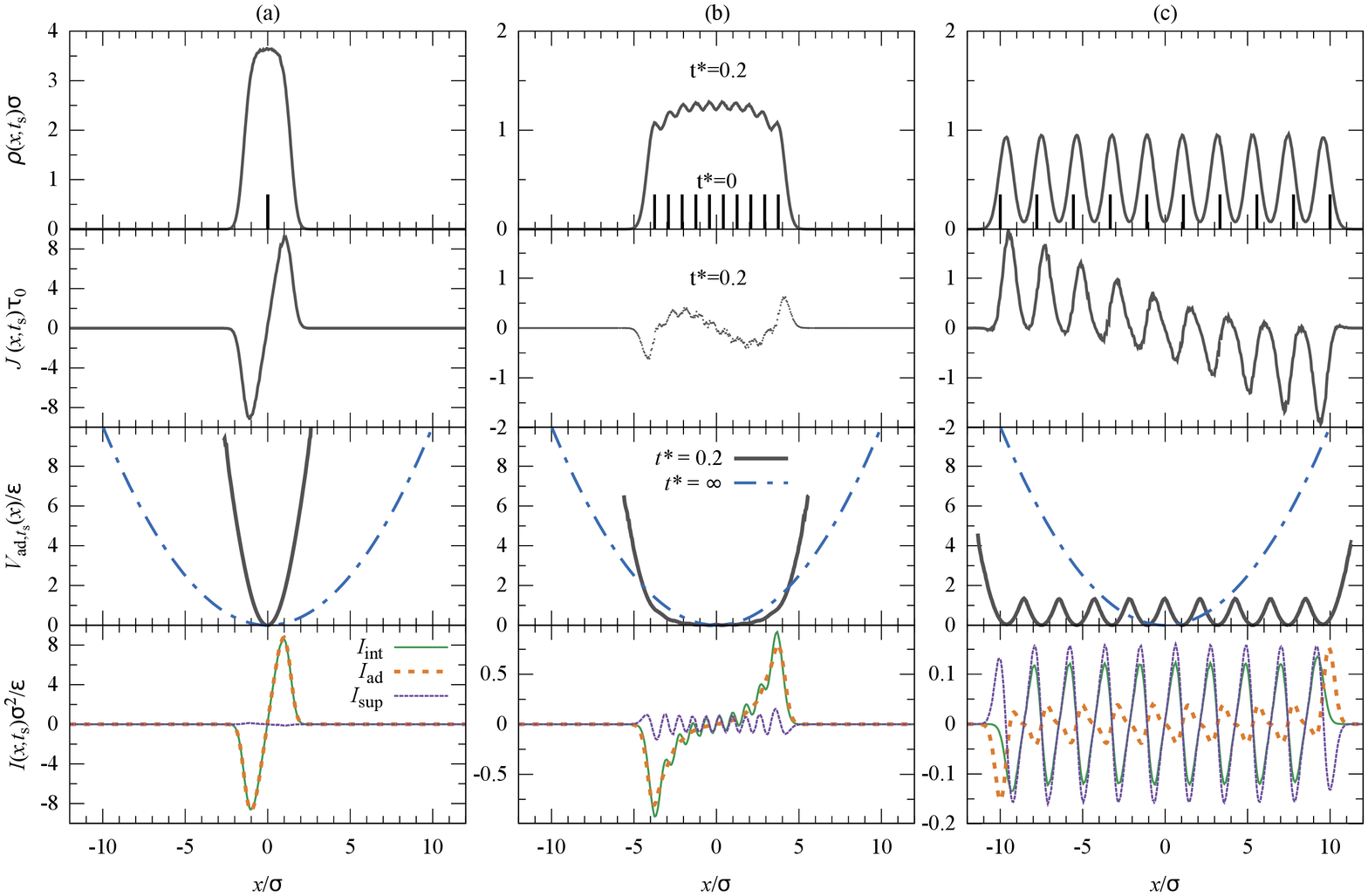}
    \caption{\label{fig:overviewcrystal}Overview analogous to Fig.~\ref{fig:overviewequi},  but for equidistant initial states at reduced temperature $T^*=0.5$ with initial nearest-neighbor separations (a) $d=0$, (b) $d=0.833\sigma$ and (c) $d=2.22\sigma$. Top panels: scaled density $\rho(x,t_\mathrm{s})\sigma$ as a function of the scaled coordinate $x/\sigma$ with initial positions indicated by vertical bars; second panels: scaled current $J(x,t_{\rm s})\tau_0$; third panels: scaled adiabatic potential $V_{\mathrm{ad},t_{\rm s}}(x)/\epsilon$. Note that the external potential that acts unchanged throughout the time evolution is equivalent to the adiabatic potential at $t^*=\infty$. Bottom panels: total internal force density $I_\mathrm{int}(x,t_{\rm s})$, adiabatic force density $I_{\mathrm{ad},t_{\rm s}}(x)$, and superadiabatic force density $I_\mathrm{sup}(x,t_s)$ in units of $\epsilon/\sigma^2$ at time $t_\mathrm{s}=\tau_0$.
}
\end{figure}

In order to examine nonequilibrium initial conditions, we chose equidistant initial positions with a distance $d$ between neighboring particles, where each position is occupied by a single particle. Note that this constitutes a nonequilibrium situation, due to the absence of multiple occupancy of the density peaks, as would occur in thermal equilibrium of the GCM.
In the case of hard rods this type of initial condition produced the largest discrepancies between total and adiabatic force densities, and thus the largest superadiabatic contributions were observed~\cite{fortini14prl}. For all times we confine the system by the external potential (\ref{eq:vext}), where $k=0.2\epsilon/\sigma^2$. The sampling time is chosen as $t_\mathrm{s}=0.2\tau_0$ and the reduced temperature is $T^*=0.5$, unless stated otherwise. Depending on the value of $d$, we can distinguish three different characteristic types of behavior for small,  intermediate and large initial separations, as shown in Fig.~\ref{fig:overviewcrystal}. In the limiting case of small separations, $d=0$, the density profile varies smoothly with distance, as depicted in the top panel of Fig.~\ref{fig:overviewcrystal}(a). At time $t_\mathrm{s}$ the density profile has expanded and the current $J(x,t_\mathrm{s})$ resembles the current for the equilibrated initial condition [as shown in the second panel of Fig.~\ref{fig:overviewcrystal}(a)]. This similarity also extends to the adiabatic potential, which is shown in the third panel of Fig.~\ref{fig:overviewcrystal}(a), as well as the splitting of the internal force density $I_\mathrm{int}(x,t_\mathrm{s})$ into $I_{\mathrm{ad},t_{\rm s}}(x)$ and $I_\mathrm{sup}(x,t_\mathrm{s})$. The superadiabatic contribution is a small correction, as shown in the bottom panel of Fig.~\ref{fig:overviewcrystal}(a), with opposite sign to $I_\mathrm{int}(x,t_\mathrm{s})$ and $I_{\mathrm{ad},t_{\rm s}}(x)$. This implies that $I_{\mathrm{ad},t_{\rm s}}(x)$ slightly overestimates $I_\mathrm{int}(x,t_\mathrm{s})$. The resemblance between the results for the current case and the above results for the initial parabolic confinement stem from the fact that the initial state with $d=0$ can be viewed as the limit of equilibrium states in $V_\mathrm{ext}(x)={kx^2/2}$ as $k\rightarrow\infty$.\par
For increased initial separations, $d=0.833\sigma$, the dynamics change significantly. While the density profile (cf.~the top panel of Fig.~\ref{fig:overviewcrystal}) still expands for $|x|>3.5\sigma$, it contracts for $|x|<3.5\sigma$. On top of this large-scale movement, local oscillations appear, as the interactions between neighboring particles become more importan; cf.~the second panel of Fig.~\ref{fig:overviewcrystal}(b). The adiabatic potential $V_{\mathrm{ad},t_{\rm s}}(x)$, shown in the third panel of Fig.~\ref{fig:overviewcrystal}(b), is considerably deformed compared to the parabolic shape of $V_\mathrm{ext}(x)$. Whereas the total internal force density $I_\mathrm{int}(x,t_\mathrm{s})$ develops oscillations, $I_{\mathrm{ad},t_{\rm s}}(x)$ remains smooth. Hence, the superadiabatic part $I_\mathrm{sup}(x,t_\mathrm{s})$ is a relevant contribution, which adds the oscillatory structure, while the adiabatic force density determines the global slope, see the bottom panel of Fig.~\ref{fig:overviewcrystal}(b).\par
\begin{figure}
  \includegraphics[width=\columnwidth]{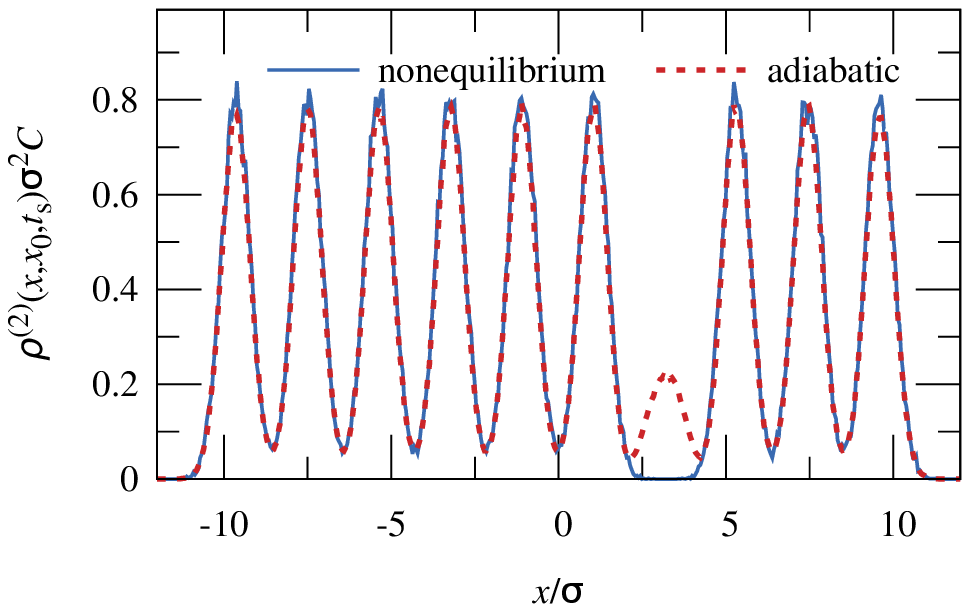}
  \caption{\label{fig:rhocs} Scaled two-body density
      $\rho^{(2)}(x,x_0,t_\mathrm{s})\sigma^2C$, where $x_0=3.17\sigma$ is a peak
      position of $\rho(x,t_\mathrm{s})$, as a function of the scaled
    distance $x/\sigma$, at time $t_\mathrm{s}=0.2\tau_0$,
    temperature $T^*=0.5$, and initial nearest neighbor separation
    distance $d=2.22\sigma$. Shown are results for the
      nonequilibrium system (blue solid line) and the corresponding
    adiabatic system (red dashed line). The normalization
      constant $C$ is chosen such that the $x$ integral of
    $\rho^{(2)}(x,x_0)$ is equal to $N-1=9$.}
\end{figure}
For large initial separations [see Fig.~\ref{fig:overviewcrystal}(c) for the case $d=2.22\sigma$], the density profile consists primarily of well-separated peaks, where the number of peaks is equal to the number of particles, $N=10$; cf.\ the top panel of Fig.~\ref{fig:overviewcrystal}(c). Since the initial density profile is wider than that for the equilibrium state in $V_\mathrm{ext}(x)$, the external force pulls the particles towards the center, hence the global slope of the current is negative; cf.\ the second panel of Fig.~\ref{fig:overviewcrystal}(c). However, nearest-neighbor interactions are clearly dominant. The adiabatic potential consists of $N$ wells, one for each density peak. The force density shown in the bottom panel of Fig.~\ref{fig:overviewcrystal}(c) possesses an oscillatory structure without any global slope. While $I_\mathrm{int}(x,t_\mathrm{s})$ and $I_{\mathrm{ad},t_{\rm s}}(x)$ oscillate in phase with each other, the superadiabatic force integral $I_\mathrm{sup}(x,t_\mathrm{s})$ is out-of-phase with both of these. This qualitative result was already reported by Fortini \emph{et al.}\ for hard rods at certain densities starting from an equidistant initial condition \cite{fortini14prl}. These authors suggested that the out-of-phase behavior of the adiabatic force density is due to contributions from microstates in which one potential well is occupied by two or more particles. In order to investigate this conjecture, we fix one of the spatial arguments of the two-body density to the position $x_0$ of one of the peaks in the one-body density $\rho(x,t_\mathrm{s})$ and evaluate the resulting function $\rho^{(2)}(x,x_0,t_\mathrm{s})$ of argument $x$. The density $\rho^{(2)}(x,x_0,t_\mathrm{s})$ is proportional to the probability of finding a particle at $x$, given that there is a particle at $x_0$, and it also appears in the integral (\ref{eq:intforcedensity}) for the force density $I_{\rm int}(x_0,t_\mathrm{s})$. In Fig.~\ref{fig:rhocs} we compare $\rho^{(2)}(x,x_0,t_\mathrm{s})$ of the nonequilibrium system to $\rho^{(2)}_{\mathrm{ad},t_\mathrm{s}}(x,x_0)$ of the adiabatic system, as a function of $x$ for $x_0=3.1\sigma$. In the nonequilibrium system the peak at $x_0$ is missing, which implies that at time $t_\mathrm{s}$ no particle has traveled from another peak to the peak at $x_0$.

Although we show only the case $x_0=3.17\sigma$, this behavior is representative and holds for all other peak positions. Therefore each of the peaks of the one-body density $\rho(x,t_\mathrm{s})$ is produced by exactly one particle, which started at $t=0$ in the initial position underneath the respective peak; see the first panel of Fig.~\ref{fig:overviewcrystal}(c). If the particle around $x_0$ is at a position on either of the wings of its own peak, it is driven back towards $x_0$ by the repulsion of the particle in the nearest neighboring peak. Hence, the total internal force density $I_\mathrm{int}(x,t_\mathrm{s})$ is positive on the left of the center of each density peak and negative on the right. However, in the adiabatic system multiple occupation of each peak is possible, as Fig.~\ref{fig:rhocs} demonstrates. If a particle is located at $x>x_0$, the small peak around $x_0$ produced by other particles exerts a force that is directed away from $x_0$. This explains the fact that the sign of $I_{\mathrm{ad},t_{\rm s}}(x)$ is opposite to that of $I_\mathrm{int}(x,t_\mathrm{s})$. As a result, $I_{\mathrm{ad},t_{\rm s}}(x)$ is out-of-phase with respect to $I_\mathrm{int}(x,t_\mathrm{s})$. Additionally, the forces from other peaks, which dominate the nonequilibrium system, are still present in the adiabatic system. The amplitude and phase of the oscillations of $I_{\mathrm{ad},t_{\rm s}}(x)$ are determined by the relation between forces from other particles in the same peak and from particles in other peaks. This mechanism also explains the increased amplitude of the last oscillations of $I_{\mathrm{ad},t_{\rm s}}(x)$ at $x=\pm10\sigma$, where forces from multiple occupation of the last peak are present, while forces in the opposite direction from a neighboring peak are absent. This larger amplitude on the outside is not observed in $I_\mathrm{int}(x,t_\mathrm{s})$.

\begin{figure}
  \includegraphics[width=\textwidth]{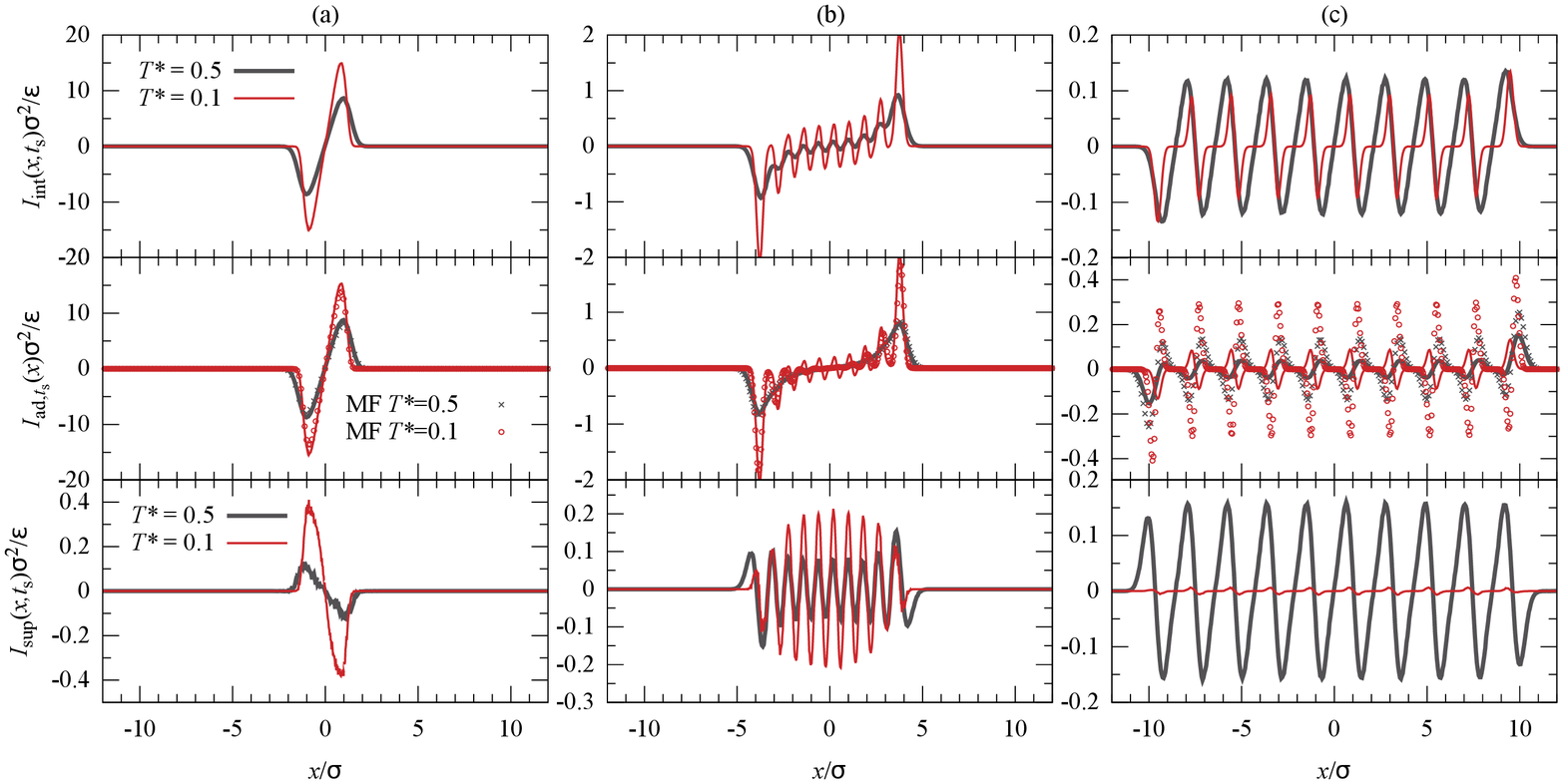}
  \caption{
\label{fig:crystalI0501l}Total internal force density $I_\mathrm{int}(x,t_{\rm s})$ (top panels), adiabatic force density $I_{\mathrm{ad},t_{\rm s}}(x)$ (middle panels) and superadiabatic force density $I_\mathrm{sup}(x,t_{\rm s})$ (bottom panels) in units of $\epsilon/\sigma^2$ {as a function of $x/\sigma$. Shown are results for} reduced temperatures $T^*=0.5$ (thick black lines) and $T^*=0.1$ (thin red lines) at $t_\mathrm{s}=\tau_0$ for equidistant initial states with initial nearest neighbor separations (a) $d=0$, (b) $d=0.833\sigma$, and (c) $d=2.22\sigma$. In each of the middle panels the adiabatic force densities $I_{\mathrm{ad},t_{\rm s}}(x)$ (lines) are also compared to the respective mean-field force density (symbols).}
\end{figure}

We next investigate the effect of decreasing the temperature to
$T^*=0.1$. We consider the force densities in the systems with the above three values for the initial separation $d$; see
Fig.~\ref{fig:crystalI0501l}. In the case of vanishing separation,
$d=0$, the total internal, the adiabatic and the superadiabatic force
density all become larger as $T^*$ is decreased; see
Fig.~\ref{fig:crystalI0501l}(a) for comparison of the results for
$T^*=0.5$ and $T^*=0.1$. The relaxation from the initial
configuration, where all particles lie on top of each other, to the
equilibrium state proceeds more slowly for the lower temperature,
which results in larger values of the internal forces at the
time $t_\mathrm{s}$, because the average distance between the
particles is smaller compared to the case of higher
temperatures. The GCM pair force reaches its maximum at distance
$\Delta x=\sigma/\sqrt{2}$, which is well below the width of the
density profile at $t_\mathrm{s}$ for $T^*=0.5$. For both temperatures
the respective superadiabatic contributions are small and the
mean-field force density provides a very good approximation to the
adiabatic and to the total force density.\par For $d=0.833\sigma$
the total internal force density $I_\mathrm{int}(x,t_\mathrm{s})$ and
the superadiabatic force density $I_\mathrm{sup}(x,t_\mathrm{s})$ at
$T^*=0.1$ are scaled versions of the respective force densities at
$T^*=0.5$; cf.\ the first and third panels of
Fig.~\ref{fig:crystalI0501l}. The adiabatic force density
$I_{\mathrm{ad},t_{\rm s}}(x)$, which does not oscillate for $T^*=0.5$, becomes
more structured when decreasing the temperature to $T^*=0.1$. It is
also apparent from the second panel of Fig.~\ref{fig:crystalI0501l}
and the magnitude of the superadiabatic force density that the
mean-field approach describes only the dynamics of the adiabatic
system well, but not that of the nonequilibrium
system. Therefore, mean-field behavior in equilibrium does not
necessarily imply mean-field behavior out-of-equilibrium for all
initial conditions.\par In the case of the large initial separation
$d=2.22\sigma$, the results at $T^*=0.5$ are qualitatively different
from those at $T^*=0.1$. For the lower temperature $I_{\mathrm{ad},t_{\rm s}}(x)$
and $I_\mathrm{int}(x,t_\mathrm{s})$ oscillate in phase with each
other and have a similar amplitude. Hence,
$I_\mathrm{sup}(x,t_\mathrm{s})$ is small. This is in accordance with
the concept of multi-occupancy generating the superadiabatic part,
since the occurrence of more than one particle in one potential well
is more strongly suppressed at lower temperatures. Furthermore, at
$T^*=0.5$ the mean-field force density is a scaled-up version of
$I_{\mathrm{ad},t_{\rm s}}(x)$. This can be understood considering that 
in mean field $\rho^{(2)}(x,x_0)=\rho(x)\rho(x_0)$. The
mean-field profile corresponding to those shown in
Fig.~\ref{fig:rhocs} therefore has a full peak at $x_0$, but
  only a smaller peak in the adiabatic profile. Hence, the mean-field
force density is in phase with $I_{\mathrm{ad},t_{\rm s}}(x)$, but has a larger
amplitude. This is not the case for $T^*=0.1$, where the
multi-occupancy mechanism is suppressed.\par
\begin{figure}
    \includegraphics[width=\columnwidth]{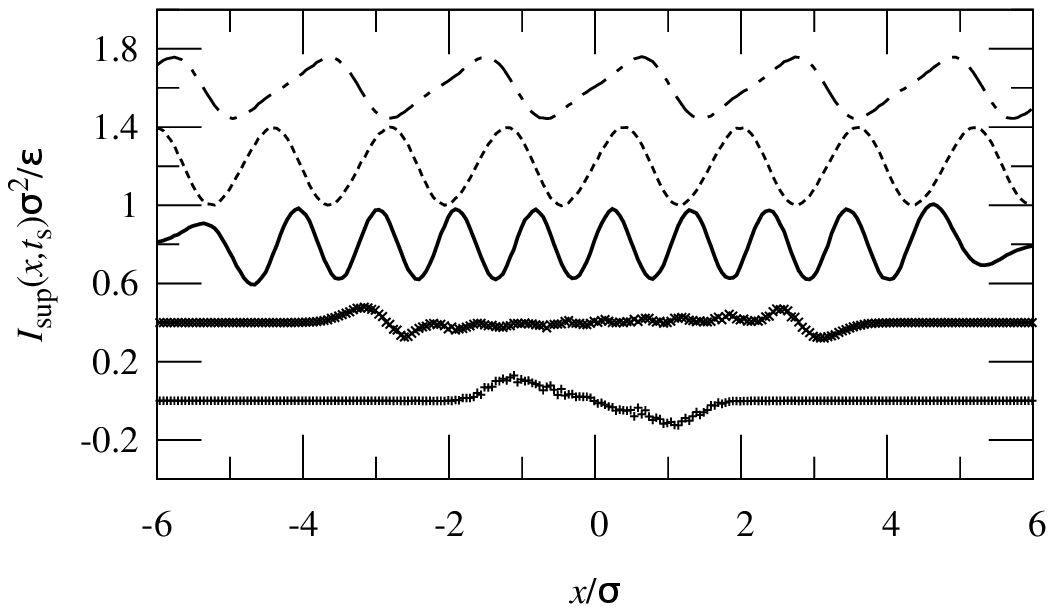}
 \caption{\label{fig:supadcry} Scaled superadiabatic force density $I_\mathrm{sup}(x,t_{\rm s})\sigma^2/\epsilon$ {as a function of $x/\sigma$} for different initial separations $d/\sigma=0$, $0.56$, $1.11$, $1.67$, $2.22$ (from bottom to top). Curves are shifted upwards by $0.4$ units for clarity. Note that the magnitude of the different results is similar.}
\end{figure}

Finally, it is remarkable that for all initial separations, ranging from $d=0$ to $d=2.22\sigma$, the superadiabatic force density $I_\mathrm{sup}(x,t_\mathrm{s})$ remains of the same order of magnitude, as shown in Fig.~\ref{fig:supadcry}. In contrast, the total internal force density $I_\mathrm{int}(x,t_\mathrm{s})$ decreases by roughly two orders of magnitude over the range of $d$ considered. This is most striking for the larger separations between $d=1.11\sigma$ and $d=2.22\sigma$, where the amplitude of the oscillations in $I_\mathrm{sup}(x,t_\mathrm{s})$ remains approximately the same. This decrease in the total force density for higher separations $d$ can be rationalized by the fact that in the nonequilibrium system the interaction is weakened by greater distance between the particles, both directly through the spatial dependence of the GCM force and indirectly, because particles take more time to come near to each other. The probability of multiple occupancy of a potential well in the adiabatic system, however, is not decreased by greater distance between the wells. Therefore, the superadiabatic force density, which corrects the additional force density from multiple occupation, stays roughly of the same magnitude.

\section{Conclusions}
\label{chap:concl}

We have investigated the splitting into adiabatic and superadiabatic contributions of the internal force density in the one-dimensional GCM for equilibrated initial states as well as for a range of nonequilibrium initial states. For this purpose we have  applied the computational scheme which was presented by Fortini \emph{et al.}\ \cite{fortini14prl} to a penetrable soft core system which differs qualitatively from the hard rod system. Thereby, we have demonstrated the generality of the approach. In particular, it is computationally feasible to solve the inverse problem of finding the adiabatic potential for a given density profile. The magnitude of the superadiabatic force density and thus the validity of the adiabatic approximation in DDFT depends strongly on the initial conditions that one considers. We have found that if the system is initially equilibrated in an external parabolic potential, then the superadiabatic contribution to the force density can be regarded as a small correction. Hence, in this case the adiabatic approximation is justified. We have found this behavior, independent of temperature in the examined temperature range.

For equidistant initial configurations we have found large superadiabatic contributions to the internal force density at high temperature. We have identified the primary source of these contributions as multiple occupancy of wells in the adiabatic potential. If multiple particles occupy the same well, they exert forces on each other which cause the adiabatic force density $I_{\mathrm{ad},t_{\rm s}}(x)$ to oscillate out-of-phase with the total internal force density $I_\mathrm{int}(x,t_{\rm s})$. This results in a large superadiabatic contribution. This mechanism is less efficient at smaller temperatures. Since multiple occupancy does not depend on the exact form of the pair potential, we expect qualitatively similar results for pair interactions other than the GCM.

Our findings are entirely consistent with those by Reinhardt and
Brader \cite{reinhardt12} who found unphysical self-interactions of
the tagged particle density fields creates an excessively fast relaxation
predicted by DDFT.

For both equilibrated initial states and small separations in the equidistant initial states we observed dynamical mean-field behavior \cite{dzubiella03}. In these cases the superadiabatic force was small. However, we also found cases where the mean-field approximation was good in equilibrium, but not in the nonequilibrium system. Hence the precise relationship between the mean-field force and the adiabatic and superadiabatic forces is complex, except in special cases. While we tried to choose conditions which allow the study of the most relevant effects, the superadiabatic part might behave unexpectedly for other initial states, which remains to be explored.  Investigation of the time evolution of superadiabatic forces might clarify the way towards a theoretical model.

Furthermore, investigating the proposed multi-occupancy mechanism on
the level of dynamical two-body correlation functions, in the
framework of the nonequilibrium Ornstein-Zernike relation
\cite{brader13noz,brader14noz}, the dynamical test-particle limit
\cite{archer07dtpl,hopkins10dtpl,brader15dtpl,stopper15pre,stopper15jcp},
or explicit many-body simulations \cite{schindler16lj} should prove
useful.

\acknowledgments We thank Daniel de las Heras for a critical reading
of the manuscript and Andrea Fortini and Joseph M. Brader for useful
discussions.


\begin{thebibliography}{10}

\bibitem{Born1928}
M. Born and V.~A. Fock, Z. Phys. {\bf 51},  165  (1928).

\bibitem{evans79}
R. Evans, Adv. Phys. {\bf 28},  143  (1979).

\bibitem{marinibettolomarconi99}
U. M. B. Marconi and P. Tarazona, J. Chem. Phys. {\bf 110},  8032
  (1999).

\bibitem{dzubiella03}
J. Dzubiella and C.~N. Likos, J. Phys.:\ Condensed Matter {\bf 15},  L147
  (2003).

\bibitem{archer04ddft}
A.~J. Archer and R. Evans, J. Chem. Phys. {\bf 121},  4246  (2004).

\bibitem{Hansen13}
J.-P. Hansen and I.~R. McDonald, {\em Theory of Simple Liquids}, 4th ed.
  (Academic Press, Amsterdam 2013).

\bibitem{royallSedimentation}
C.~P. Royall, J. Dzubiella, M. Schmidt, and A. {van Blaaderen}, Phys. Rev.
  Lett. {\bf 98},  188304  (2007).
  \bibitem{evans16pc}
  For an overview of current work, see the preface for the Special Issue on "New developments in classical density functional theory" by R. Evans, M. Oettel, R. Roth and G. Kahl, J. Phys.:\ Condensed Matter {\bf 28}, 240401 (2016).

\bibitem{fortini14prl}
A. Fortini, D. de~las Heras, J.~M. Brader, and M. Schmidt, Phys. Rev. Lett.
  {\bf 113},  167801  (2014).

\bibitem{penna06}
F. Penna and P. Tarazona, J. Chem. Phys. {\bf 124},  164903  (2006).

\bibitem{reinhardt12}
J. Reinhardt and J.~M. Brader, Phys. Rev. E {\bf 85},  011404  (2012).

\bibitem{stillinger76}
F.~H. Stillinger, J. Chem. Phys. {\bf 65},  3968  (1976).

\bibitem{brader01oned}
J.~M. Brader and R. Evans, Physica A {\bf 306},  287  (2002).

\bibitem{santos07oned}
A. Santos, Phys. Rev. E {\bf 76},  062201  (2007).

\bibitem{lutz04}
C. Lutz, M. Kollmann, and C. Bechinger, Phys. Rev. Lett. {\bf 93},  026001
  (2004).


\bibitem{schmidt13pft}
M. Schmidt and J.~M. Brader, J. Chem. Phys. {\bf 138},  214101  (2013).

\bibitem{delasheras16pc}
D. de~las Heras, J.~M. Brader, A. Fortini, and M. Schmidt, 
J. Phys.:\ Condensed Matter {\bf 28}, 244024 (2016).

\bibitem{delasheras14prl}
D. de~las Heras and M. Schmidt, Phys. Rev. Lett. {\bf 113},  238304  (2014).


\bibitem{louis00}
A.~A. Louis, P.~G. Bolhuis, J.~P. Hansen, and E.~J. Meijer, Phys. Rev. Lett.
  {\bf 85},  2522  (2000).

\bibitem{archer01}
A.~J. Archer and R. Evans, Phys. Rev. E {\bf 64},  041501  (2001).

\bibitem{archer02}
A.~J. Archer and R. Evans, J. Phys.:\ Condensed Matter {\bf 14},  1131  (2002).

\bibitem{archer02el}
A.~J. Archer, R. Evans, and R. Roth, Europhys. Lett. {\bf 59},  526  (2002).

\bibitem{archer03}
A.~J. Archer and R. Evans, J. Chem. Phys. {\bf 118},  9726  (2003).

\bibitem{archer04}
A.~J. Archer, C.~N. Likos, and R. Evans, J. Phys.:\ Condensed Matter {\bf 16},
  L297  (2004).

\bibitem{brader13noz}
J.~M. Brader and M. Schmidt, J. Chem. Phys. {\bf 139},  104108  (2013).

\bibitem{brader14noz}
J.~M. Brader and M. Schmidt, J. Chem. Phys. {\bf 140},  034104  (2014).

\bibitem{archer07dtpl}
A.~J. Archer, P. Hopkins, and M. Schmidt, Phys. Rev. E {\bf 75},  040501(R)
  (2007).

\bibitem{hopkins10dtpl}
P. Hopkins, A. Fortini, A.~J. Archer, and M. Schmidt, J. Chem. Phys. {\bf 133},
   224505  (2010).

\bibitem{brader15dtpl}
J.~M. Brader and M. Schmidt, J. Phys.:\ Condensed Matter {\bf 27},  194106
  (2015).

\bibitem{stopper15pre}
D. Stopper, K. Marolt, R. Roth, and H. Hansen-Goos, Phys. Rev. E {\bf 92},
  022151  (2015).

\bibitem{stopper15jcp}
D. Stopper, R. Roth, and H. Hansen-Goos, J. Chem. Phys. {\bf 143},  181105
  (2015).

\bibitem{schindler16lj}
T. Schindler and M. Schmidt, to be published.


\end{thebibliography}

\end{document}